\documentclass[conference,compsoc]{IEEEtran}
  \usepackage{cite}
\usepackage{hyperref}
\usepackage{xurl}

\ifCLASSINFOpdf
\else
\fi

\usepackage{datetime2}
\usepackage{xcolor,bm} 
\usepackage{enumerate,enumitem} 

\usepackage{booktabs,multirow} 
\usepackage{amssymb}
\usepackage{amsmath,amsfonts,amsthm, mathtools} 

\usepackage{color}
\usepackage{tikz} 
\usepackage[font=small,labelfont=bf]{caption}
\usepackage{subcaption} 
\usepackage{quantikz}
\usepackage{adjustbox}
\usepackage{float}

\newcommand{\beqa} {\begin{eqnarray}}
\newcommand{\eeqa} {\end{eqnarray}}

\newcommand{\no}{\nonumber}

\newcommand{\bi}{\begin{itemize}}
\newcommand{\ei}{\end{itemize}}

\newcommand{\bftab}{\fontseries{b}\selectfont}
\begin{document}
%
\title{A Novel Image Classification Framework Based on Variational Quantum Algorithms}


\author{\IEEEauthorblockN{Yixiong Chen}
\IEEEauthorblockA{Email: chenyixiong516@msn.com}
}


%


\maketitle

\begin{abstract}
Image classification is a crucial task in machine learning with widespread practical applications. The existing classical framework for image classification typically utilizes a global pooling operation at the end of the network to reduce computational complexity and mitigate overfitting. However, this operation often results in a significant loss of information, which can affect the performance of classification models. To overcome this limitation, we introduce a novel image classification framework that leverages variational quantum algorithms (VQAs)—hybrid approaches combining quantum and classical computing paradigms within quantum machine learning.  The major advantage of our framework is the elimination of the need for the global pooling operation at the end of the network. In this way, our approach preserves more discriminative features and fine-grained details in the images, which enhances classification performance. Additionally, employing VQAs enables our framework to have fewer parameters than the classical framework, even in the absence of global pooling, which makes it more advantageous in preventing overfitting. We apply our method to different state-of-the-art image classification models and demonstrate the superiority of the proposed quantum architecture over its classical counterpart through a series of experiments on public datasets. Our experiments show that the proposed quantum framework achieves up to a 9.21\% increase in accuracy and up to a 15.79\% improvement in F1 score, compared to the classical framework.
\end{abstract}


\section{Introduction}\label{introduction}
In light of recent breakthroughs in quantum technologies, particularly the availability of noisy intermediate-scale quantum (NISQ) processors \cite{cheng2023noisy}, the field of quantum machine learning has attracted growing concerns and triggered an enormous amount of work \cite{schuld2015introduction, biamonte2017quantum, dunjko2018machine, zhang2020recent, cerezo2022challenges}. Quantum machine learning is an interdisciplinary field that combines concepts from quantum physics and machine learning to develop innovative algorithms and models. By harnessing the power of quantum computing, quantum machine learning aims to improve the performance of machine learning algorithms. While still an emerging discipline, quantum machine learning has already demonstrated promising quantum extensions to classical machine learning techniques including support vector machines \cite{varatharajan2018big}, clustering \cite{kerenidis2018q, otterbach2017unsupervised},  and principal component analysis \cite{article}.

Image classification is a fundamental task in computer vision that involves categorizing images into predefined classes or categories. Over the past few years, this field has witnessed rapid progress. Convolutional neural networks (CNNs) emerged as a breakthrough technique, significantly improving image classification accuracy. Various CNN architectures \cite{krizhevsky2012imagenet, lin2013network, simonyan2014very, szegedy2015going, he2016deep, huang2017densely, tan2019efficientnet} have progressively advanced state-of-the-art (SOTA) results on benchmark datasets such as ImageNet \cite{deng2009imagenet}. Meanwhile, self-attention models like Transformers \cite{vaswani2017attention} in natural language processing have been introduced to computer vision. The corresponding transformer-based vision models, such as Vision Transformer (ViT) \cite{dosovitskiy2010image} and Swin Transformer \cite{liu2021swin}, can achieve compelling results on image recognition and even outperform CNNs on some tasks. More recently, hybrid approaches combining CNN and transformer modules are also gaining attention. These hybrid CNN-transformer models, including CoAtNet \cite{dai2021coatnet} and MaxViT \cite{tu2022maxvit}, have been demonstrated to improve the generalizability of pure transformer-based models and achieve excellent performance in image classification without  being pre-trained on large-scale image datasets.

In most image classification architectures, the input images are processed by a backbone model, also referred to as a feature extractor, which extracts relevant features from the input. This backbone model generally consists of a series of convolutional layers, transformer-based layers, or a combination of both. The resulting feature maps are typically fed into a global pooling layer followed by fully connected layers and classifier. Global pooling is a commonly employed technique in the field of image recognition \cite{lin2013network, simonyan2014very, szegedy2015going, he2016deep, huang2017densely, tan2019efficientnet, liu2021swin, dai2021coatnet, tu2022maxvit}. It  aggregates the spatial information of feature maps across the entire image into a single vector representation. The advantage of global pooling lies in its ability to capture the overall context and semantic information of an image, resulting in a compact and fixed-length representation that can be easily fed into subsequent layers for classification tasks. Global pooling also helps to mitigate the spatial variance of features, enabling the model to achieve translation invariance and robustness to object deformations. Moreover, it reduces the dimensionality of the feature maps, leading to computational efficiency and alleviating the risk of overfitting.

While global pooling provides some benefits for image classification, it has several limitations. Collapsing an entire feature map into a single vector results in the complete loss of fine-grained spatial information. The compact fixed-length representation has also restricted expressive capacity compared to a fully-connected layer operating on spatial data, which may hinder capturing complex spatial relationships. This can be disadvantageous for tasks that require precise object localization, discriminative information or detailed spatial information. Furthermore, uninformative features corresponding to background or irrelevant regions can dominate the global summary statistic and render it unresponsive to salient features. Additionally, since global pooling treats the entire image equally, it may not effectively capture local variations or small-scale patterns that are important for accurate recognition. 

Researchers have explored different strategies to enhance the performance of global pooling and mitigate its drawbacks. Firstly, several variants of global pooling are proposed to provide a trade-off between the global average pooling and global max pooling with a nonlinear log-average-exp (LAE) \cite{zhang2018alphamex, lowe2021logavgexp} or log-sum-exp (LSE) \cite{sun2016pronet} function, which helps extract features more effectively.  Another approach is to leverage pyramid pooling \cite{he2015spatial, jose2018pyramid, qi2018concentric, lin2019fourier} as an alternative to global pooling. Pyramid pooling methods divide the image into multiple regions of different scales and extracts features from each region independently. This allows for the capture of multi-scale information and the preservation of spatial details. Moreover, attention mechanisms have also been integrated with global pooling to improve its discriminative power \cite{qiu2018global, zhang2020global, behera2021context}. These mechanisms selectively emphasize salient regions or features while suppressing irrelevant ones, enhancing the model's ability to capture important information. In addition, spatial transformer networks (STNs) \cite{jaderberg2015spatial} have been proposed to enable adaptive spatial transformations prior to pooling. STNs apply learned transformations to align features, enhancing the localization accuracy of global pooling.  Furthermore, the method of global second-order pooling \cite{li2018towards, gao2019global, wang2020deep, song2022eigenvalues} is also exploited for capturing more discriminative image representations. This approach calculates covariance matrices to obtain higher-order statistical information between features.

 In contrast to the above methods, our work employs quantum machine learning algorithms to address the issue of global pooling for the task of image classification. Quantum machine learning has the potential to revolutionize the area of image
classification as quantum algorithms can process large datasets of images more efficiently than classical algorithms, leading to faster and more accurate classification  \cite{dang2018image, kerenidis2019quantum, mari2020transfer, henderson2020quanvolutional, li2020quantum, liu2021hybrid, henderson2021methods, houssein2021hybrid, sagingalieva2022hyperparameter, hur2022quantum, senokosov2023quantum, melnikov2023quantum}. In this paper, we propose a novel image classification framework based on variational quantum algorithms (VQAs) \cite{mcclean2016theory, bharti2022noisy, cerezo2021variational}  in the context of quantum machine learning. VQAs are a class of quantum algorithms that utilize classical computers to optimize a parameterized quantum circuit. Due to their hybrid quantum-classical approach that has potential noise resilience \cite{mcclean2016theory}, VQAs have been considered as one of the leading candidates to achieve application-oriented quantum computational advantage on NISQ devices.   

The proposed framework differs from classical architectures primarily in that global pooling  operations at the end of the network are no longer required. This is due to the introduction of the variational quantum circuits (VQCs) \cite{cerezo2021variational, mitarai2018quantum, farhi2018classification} which possess remarkable data storage capacity. Consequently, our framework can fully utilize the feature maps extracted via the backbone model for image classification, which is particularly advantageous for tasks requiring fine-grained features. Moreover, despite the removal of the global pooling layer,  our framework does not suffer from the issue of excessive model parameters and overfitting, as commonly encountered in classical deep learning frameworks. In contrast, our framework actually has fewer parameters, thanks to the robust data storage and expressive capabilities of the variational quantum circuits. Furthermore, our framework exhibits high flexibility and adaptability, making it applicable to most of existing deep learning models for image classification.

In summary, the contributions of our work are
\bi
\item We propose a novel image classification framework based on variational quantum algorithms, which does not require global pooling before the classification layer and enables models to capture more discriminative details from images using fewer parameters. To the best of our knowledge, our work is the first attempt to tackle the problem of global pooling in the classical image classification framework by employing quantum machine learning algorithms.
\item We evaluate our proposed framework on four challenging datasets, namely Croatian Fish \cite{jager2015croatian}, Aircraft \cite{maji2013fine}, Breast Ultrasound image \cite{al2020dataset}, and Apples or Tomatoes \cite{AOT}. Experimental results demonstrate that our framework significantly outperform the classical framework.
\ei

\section{Method}\label{method}

\subsection{Preliminaries}

\subsubsection{Global Pooling} 
\begin{figure}[h!]
\centering
\includegraphics[width=0.4\textwidth]{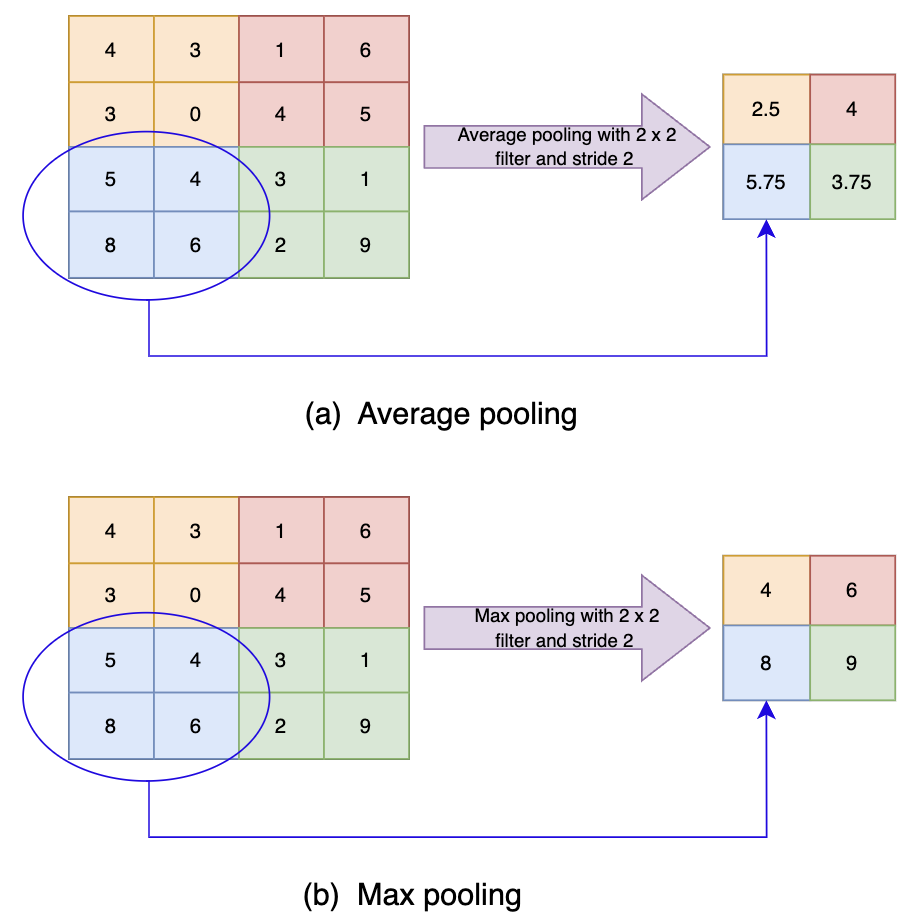}\vspace{0.2cm}
\caption{Examples of two types of pooling operations. (a) Average pooling with a pooling area of size 2x2 and stride of 2. (b) Max pooling with a pooling area of size 2x2 and stride of 2.}
\label{pooling}
\end{figure}

Pooling is a common operation in image classification models that plays a vital role in reducing spatial dimensions and capturing invariant  features. It aims to down sampling feature maps by summarizing the presence of features in patches of the feature map. Two commonly used pooling methods are average pooling and max pooling. Average pooling calculates the average value of the feature within each pooling region while max pooling selects the maximum value of the feature. Examples of these two pooling approaches are shown in Fig. \ref{pooling}.

\begin{figure}[h!]
\centering
\includegraphics[width=0.34\textwidth]{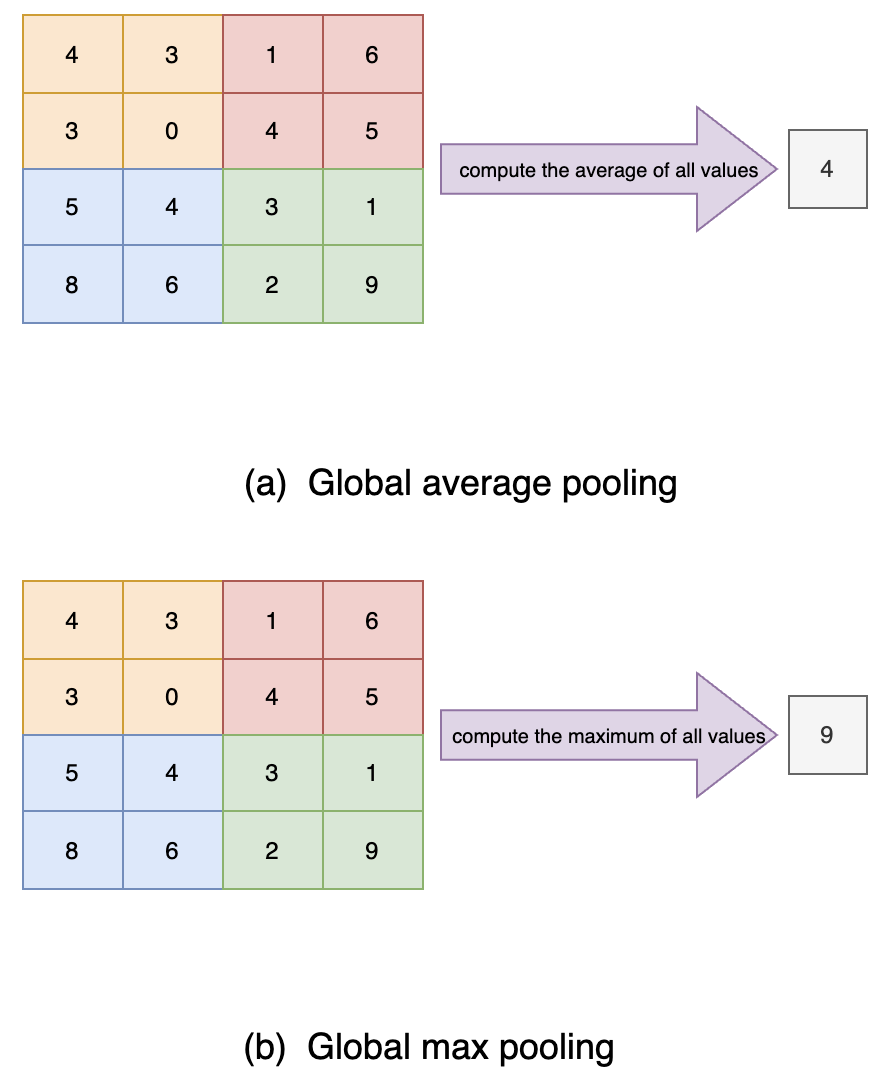}\vspace{0.2cm}
\caption{Examples of two types of global pooling operations. (a) Global average pooling. (b) Global max pooling.}
\label{GP}
\end{figure}

Global Pooling is a variant of pooling operation where the kernel size is equal to the size of the input feature map. Essentially, it computes a single value for each feature channel by summarizing the entire feature map. The two most common types of global pooling are global average pooling (GAP) and global max pooling (GMP), as illustrated in Fig. \ref{GP}, which compute the average value and maximum value of each feature channel respectively. Global pooling is often used as a bridge between the backbone model and  fully connected layers in classical image classification architectures. They help to reduce overfitting by minimizing the total number of parameters in the model. However, they can also lead to substantial loss of spatial information and critical details by transforming an entire feature map to a single value. This prevents the model from capturing local details effectively and thus limits the model performance.

\subsubsection{Quantum Basics} 
Here, we provide a brief introduction to some fundamental concepts of quantum computing that are essential for understanding this paper.
\bi
\item \textbf{Qubit and Quantum State}:
The qubit is the basic unit of information in quantum computing. Unlike a classical bit which has a value of either 0 or 1, a qubit can exist in a combination of the two states. This is called superposition. The quantum state of a qubit is described by a vector in a two-dimensional Hilbert space, commonly represented as 
\beqa
|\psi\rangle = \alpha |0\rangle + \beta |1\rangle  \label{general},
\eeqa
where $|0\rangle$ and $|1\rangle$ are  two computational basis states,  and $\alpha$ and $\beta$ are complex  probability  amplitude corresponding to each basis state, satisfying 
\beqa
|\alpha|^2 + |\beta|^2 = 1. \no
\eeqa

\item \textbf{Quantum gate and Quantum circuit}:  Quantum gates are unitary operators or equivalently unitary matrices which are applied to states of qubits and perform unitary transformation. Common single-qubit quantum gates include the Hadamard gate, Pauli gates and their corresponding rotation gates. In particular, single-qubit rotation gates are parametric gates that depend on one parameter and allow for a more flexible range of operations. These rotation gates, denoted by $R_i(\theta)$, perform a rotation around the $i$ axis on the Bloch sphere \cite{bloch1946nuclear} by an angle $\theta$, where $i \in \{x,y,z\}$ and $\theta \in [0,2\pi)$. Other examples of common gates include two-qubit gates such as the CNOT gate which can induce quantum entanglement between qubits. A summary of these quantum gates can be found in Table \ref{gates}.
\begin{table}[h!]
\centering
\begin{adjustbox}{scale=0.55}
\large
\begin{tabular}{lcc}
\toprule
\textbf{Quantum Gate} & \textbf{Matrix Expression} & \textbf{Symbol} \\
\midrule
Pauli-X Gate  & $\begin{bmatrix} 0 & 1 \\ 1 & 0 \end{bmatrix}$ & \begin{tikzcd}[column sep=0.7cm,row sep=0.7cm] \qw & \gate{X} & \qw \end{tikzcd} \\
\midrule
Pauli-Y Gate & $\begin{bmatrix} 0 & -i \\ i & 0 \end{bmatrix}$ & \begin{tikzcd}[column sep=0.7cm,row sep=0.7cm] \qw & \gate{Y} & \qw \end{tikzcd} \\
\midrule
Pauli-Z Gate & $\begin{bmatrix} 1 & 0 \\ 0 & -1 \end{bmatrix}$ & \begin{tikzcd}[column sep=0.7cm,row sep=0.7cm] \qw & \gate{Z} & \qw \end{tikzcd} \\
\midrule
$R_x$ Gate & $\begin{bmatrix} \cos(\theta/2) & -i\sin(\theta/2) \\ -i\sin(\theta/2) & \cos(\theta/2) \end{bmatrix}$ & \begin{tikzcd}[column sep=0.7cm,row sep=0.7cm] \qw & \gate{R_x(\theta)} & \qw \end{tikzcd} \\
\midrule
$R_y$ Gate & $\begin{bmatrix} \cos(\theta/2) & -\sin(\theta/2) \\ \sin(\theta/2) & \cos(\theta/2) \end{bmatrix}$ & \begin{tikzcd}[column sep=0.7cm,row sep=0.7cm] \qw & \gate{R_y(\theta)} & \qw \end{tikzcd} \\
\midrule
$R_z$ Gate & $\begin{bmatrix} e^{-i\theta/2} & 0 \\ 0 & e^{i\theta/2} \end{bmatrix}$ & \begin{tikzcd}[column sep=0.7cm,row sep=0.7cm] \qw & \gate{R_z(\theta)} & \qw \end{tikzcd} \\
\midrule
Hadamard Gate & $\frac{1}{\sqrt{2}}\begin{bmatrix} 1 & 1 \\ 1 & -1 \end{bmatrix}$ & \begin{tikzcd}[column sep=0.7cm,row sep=0.7cm] \qw & \gate{H} & \qw \end{tikzcd} \\
\midrule
CNOT Gate & $\begin{bmatrix} 1 & 0 & 0 & 0 \\ 0 & 1 & 0 & 0 \\ 0 & 0 & 0 & 1 \\ 0 & 0 & 1 & 0 \end{bmatrix}$ & \begin{tikzcd}[column sep=0.7cm,row sep=0.7cm] \qw & \ctrl{1} & \qw \\ \qw & \targ{}& \qw \end{tikzcd} \\
\bottomrule
\end{tabular}
\end{adjustbox}
\captionsetup{ width=0.8\linewidth}
\caption{Summary of common single-qubit and two-qubit quantum gates.}
\label{gates}
\end{table}
A quantum circuit is a collection of interconnected quantum gates and it is constructed by arranging the gates in a specific order to perform desired computations or quantum algorithms. Quantum gates and circuits play a crucial role in quantum computing, enabling the manipulation and processing of quantum information. 
\item \textbf{Observable and Measurement}: Observables in quantum physics are self-adjoint operators acting on quantum states. The eigenvalues of observables are real numbers and represent physical quantities (e.g. position, energy and momentum) that can be measured. A measurement is the testing or manipulation of a physical system in order to yield a numerical result of the observable. A qubit can be in a superposition state (i.e. combination of all possible basis states) before the measurement. But after the measurement, it will collapse to one of those basis states with the probability obtained as the norm square of the corresponding probability amplitude (e.g. $|\alpha|^2$ for the state $|0\rangle$ in \eqref{general}). Therefore, measurement results in quantum physics are probabilistic in general.

\ei

\subsection{Variational Quantum Algorithms} 
Variational Quantum Algorithms (VQAs) are a class of quantum algorithms that leverage the strengths of both classical and quantum computing to solve computational problems. Unlike traditional quantum algorithms that rely on precise and exact quantum operations, VQAs make use of variational quantum circuits, which provide a more flexible and adaptable approach to quantum computation, particularly in the context of noisy or imperfect quantum systems. A variational quantum circuit is comprised of three components.
\begin{itemize}
\item \textbf{Encoding module}. \,The encoding module is responsible for mapping classical data into a quantum state representation. It encodes the classical input into a quantum state that can be effectively manipulated and processed by the subsequent modules. Examples of quantum encoding methods include amplitude encoding, angle encoding, and basis encoding. A comprehensive review of these methods can be found in \cite{dataencoding}. In the current work, we focus on amplitude encoding which will be introduced in detail in subsection \ref{QAE}. Let us denote by $E(x)$ the encoding operator where $x$ is the input vector. Then the encoded quantum state is obtained as
\beqa
|x\rangle = E(x) |0\rangle.
\eeqa

\item \textbf{Parameterized module}.\, Following the encoding process, the parameterized module applies a series of single- and multi-qubit gates to the encoded quantum state.  Single-qubit gates are primarily parametric rotation gates (e.g.  $R_x$ gate). Multi-qubit gates, usually CNOT gates or parametric controlled rotation gates, are used to generate correlated or entangled quantum states. This combination of single- and multi-qubit gates collectively form a parameterized layer in the quantum circuit designed to extract task-specific features. This layer can be repeated multiple times to extend the feature space. In this sense, the parameterized module is implemented by a parameterized quantum circuit which is also referred to as an ansatz in the context of quantum computing. Let us denote all unitary operations within the parameterized module as  $U(\theta)$, the resulting quantum state will be
\begin{equation}
	|x,\theta\rangle = U(\theta) |x\rangle
\end{equation}
where $\theta$ represent all trainable parameters in the module. \\
\item \textbf{Decoding module}.\, Finally, the decoding module extracts useful information from the final state of the quantum system. To be more specific, this module transforms the final quantum state into a classical vector $f(x,\theta)$ by using a map:
\beqa
\mathcal{M}: \quad  |x,\theta\rangle \rightarrow f(x,\theta).
\eeqa
This classical vector $f(x,\theta)$ is actually the expectation value of certain local observables $A^{\otimes m}$ (e.g. Pauli-Z operator $\sigma_z^{\otimes m}$) obtained from repeated measurements
\beqa
\label{expv}
f(x,\theta) = \langle x,\theta| A^{\otimes m} |x,\theta\rangle,
\eeqa
where $\otimes$ denotes the tensor product operation of quantum operators and $m$ indicates the number of qubits the operator $A$ acts on. Here, $m$ is equal or smaller than the total number of qubits $n$ in the quantum system. The classical vector $f(x,\theta)$ can be used as the input features for the subsequent layer in the model.
\end{itemize}
The structure of the variational quantum circuit is illustrated in Fig. \ref{vqc}. 
\begin{figure}
    \centering
    \begin{tikzpicture}
        \node[scale=0.7] {
        \begin{quantikz}
            \ket{0} & \gate[4]{E(x)} \gategroup[4,steps=1,style={dashed,
            rounded corners,fill=blue!0, inner xsep=2pt},
            background,label style={label position=below,anchor=
            north,yshift=-0.4cm}]{{\small Encoding }} & \qw & \gate[4][4cm]{U(\theta)} \gategroup[4,steps=1,style={dashed,
            rounded corners,fill=blue!0, inner xsep=2pt},
            background,label style={label position=below,anchor=
            north,yshift=-0.4cm}]{{\small Parameterized quantum circuit }} & \qw&\meter{} \gategroup[4,steps=1,style={dashed,
            rounded corners,fill=blue!0, inner xsep=2pt},
            background,label style={label position=below,anchor=
            north,yshift=-0.4cm}]{{\small Decoding }} & \qw \\
            \ket{0} &&\qw&&\qw&\meter{} & \qw \\
            \ket{0} &&\qw&&\qw&\meter{} & \qw \\
            \ket{0} &&\qw&&\qw&\meter{} & \qw
        \end{quantikz}
        };
    \end{tikzpicture}
    \captionsetup{width=0.8\linewidth}
    \caption{An example of a variational quantum circuit. The classical input data $x$ is encoded into a quantum state by the encoding module $E(x)$. This encoded state is then transformed by the following parameterized quantum circuit $U(\theta)$, namely the parameterized module, where $\theta$ are learnable parameters. Finally, the decoding module extracts classical information from the final quantum state by performing quantum measurements. }
    \label{vqc}
\end{figure}

VQAs employ a hybrid quantum-classical approach for optimization. The variational quantum circuit is executed on a quantum computer to generate measurement outcomes, while classical optimization techniques, such as stochastic gradient descent or Adam, are employed to update the parameters of the circuit.  In this sense, variational quantum circuits can be seen as quantum
analogs of classical neural networks, and thus can easily be
used for various machine learning tasks, such as classification and regression. It has been pointed out that variational quantum circuits are more expressive than classical neural networks \cite{henderson2020quanvolutional, mitarai2018quantum, havlivcek2019supervised}.
\subsection{Amplitude Embedding} \label{QAE}
Amplitude encoding is a technique used in quantum computing to encode classical data into the amplitudes of quantum states.  To be more specific,  it encodes a normalized classical  
$N$-dimensional vector $x$ into amplitudes of an $n$-qubit quantum state with $n=\lceil \log_2 N \rceil$:
\beqa
|x\rangle = \sum_i^N x_i |i\rangle 
\eeqa
where $|x\rangle$ is the encoded state and $|i\rangle $ is the $i$-th computational basis state. The classical vector $x$ must satisfy normalization condition: $|x|^2=1$, as it represents the amplitudes of a quantum state. 

Amplitude encoding is a particularly efficient data encoding scheme as it enables a quantum computer to process a large amount of data simultaneously due to the principles of superposition and interference. One common use of  amplitude encoding is in quantum machine learning algorithms, where high-dimensional data vectors are encoded into the amplitudes of a quantum state. This allows quantum algorithms to operate on high-dimensional data in ways that could potentially be more efficient than classical algorithms.

\subsection{Image Classification Framework Based on Variational Quantum Algorithms}  
\begin{figure}[h!]
\centering
\includegraphics[width=0.4\textwidth]{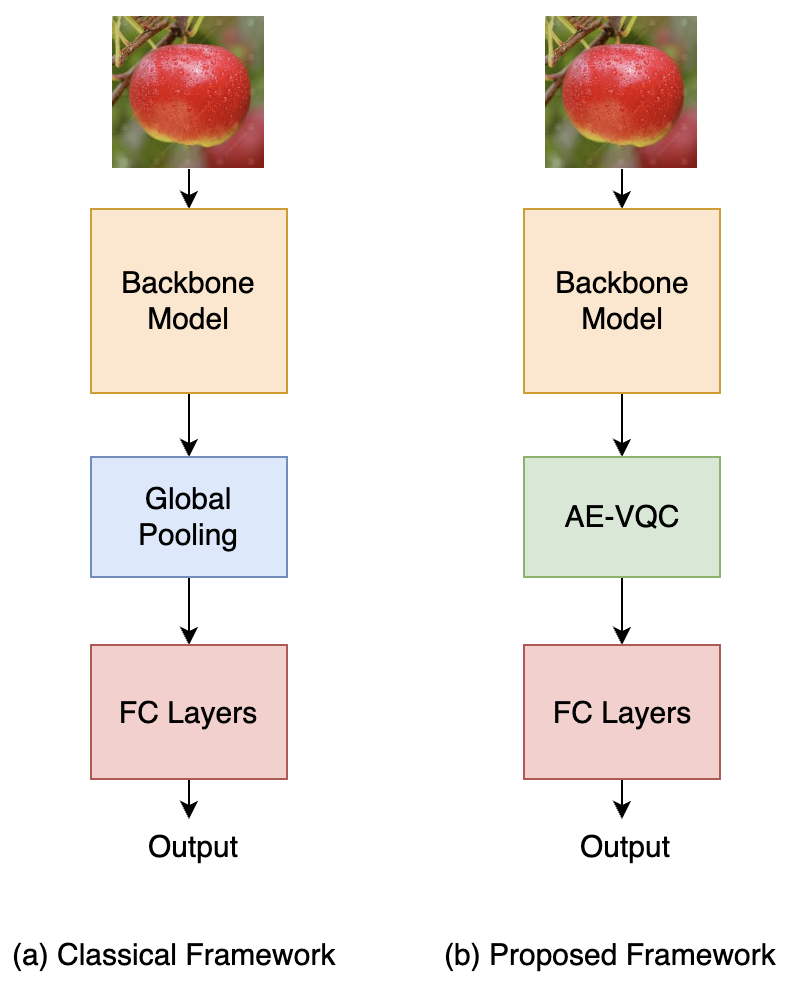}\vspace{0.2cm}
\caption{Comparison of two image classification frameworks. (a) Classical framework. The input images are transformed by a backbone model into relevant features. The following global pooling module downsample these feature maps and outputs a fix-length vector which is fed into fully connected layers for the final classification. (b) Proposed framework. The global pooling module is replaced by a variational quantum circuit (VQC) with amplitude encoding which we denote by AE-VQC. The feature maps extracted by the backbone model are directly fed into this AE-VQC without dimensionality reduction. The outputs of the AE-VQC are transformed into classification results via fully connected layers. }
\label{framework}
\end{figure}

The classical image classification framework, as shown in Fig. \ref{framework}(a),  generally consists of a backbone model, a global pooling layer and several fully connected layers. Specifically, the backbone model is applied to input images to extract relevant features. The following global pooling module is then used to reduce the spatial dimensions of these feature maps. Finally, fully connected layers learn higher-level representations from these pooled features and perform the classification. As previously mentioned, the global pooling operation results in a significant loss of information, potentially degrading model performance. To address this problem, we propose in this paper a novel image classification framework, as depicted in Fig. \ref{framework}(b). This framework is based on variational quantum algorithms and it is a departure from the classical framework. The primary idea of this framework is to eliminate the global pooling operation commonly used at the end of the classical framework and replace it by a variational quantum circuit with amplitude encoding which we refer to as AE-VQC. The other parts of the proposed framework remain the same as the classical framework. 

Next, we introduce the details of the AE-VQC in our framework. Due to the absence of the global pooling layer following the backbone model, the input size of the variational quantum circuit becomes extremely large. For example, in the ResNet-18 model, the size of the feature maps after the last convolutional layer is $512 \times 7 \times 7 = 25088$. Therefore, amplitude encoding is selected in the variational quantum circuit to effectively store a large amount of information with fewer qubits. In the case of ResNet-18, only $\lceil \log_2 25088 \rceil  = 15$ qubits are needed to encode those feature maps. If we were to use other data encoding methods, the variational quantum circuit would require a much larger number of qubits (e.g. 25088 qubits required by angle encoding for the case of ResNet-18), which is impractical in the current NISQ era. After the classical information is encoded into an initial quantum state, an ansatz (i.e. a parameterized quantum circuit) is used to transform this quantum state. The design of the ansatz is arbitrary. Here, we consider two ansatzes: Ansatz 1 and Ansatz 2, as shown in Fig. \ref{ansatzes}. Ansatz 1 simply applies successive parameterized $R_x$ gates on each qubit. Ansatz 2 transforms the initial state into a uniform superposition state by using the H gate, followed by repeatedly applying a module composed of $R_x$, $R_z$ and CNOT gates. With the introduction of the CNOT gate, Ansatz 2 may create quantum entanglement. Finally, in the decoding layer, we measure the Pauli-$Z$ operators and use their expectation values as outputs, which are then passed into the subsequent classical layers.
\begin{figure}[h!]
  \centering
  \begin{subfigure}[b]{0.3\textwidth}
    \centering
    \scalebox{0.8}{
    \begin{quantikz}[row sep=0.3cm, column sep=0.3cm]
    \qw&  \qw & \gate{R_x(\theta_1)} \gategroup[3,steps=1,style={dashed,rounded
corners,fill=blue!0, inner
xsep=2pt},background,label style={label
position=below right,anchor=north,xshift=0.4cm,yshift=-0.6cm}]{{$\times$ D}} & \qw &  \qw\\
    \qw& \qw & \gate{R_x(\theta_2)} & \qw &  \qw\\
   \qw&  \qw & \gate{R_x(\theta_3)} & \qw &  \qw
\end{quantikz}}
    \caption{Ansatz 1}
    \label{fig:circuitA}
  \end{subfigure}\vspace{0.5cm} 
 \vfill
  \begin{subfigure}[b]{0.3\textwidth}
    \centering
     \scalebox{0.8}{
\begin{quantikz}
                \qw   &  \gate{H} & \gate{R_x(\theta_1)}  \gategroup[3,steps=4,style={dashed,rounded
corners,fill=blue!0, inner
xsep=2pt},background,label style={label
position=below right,anchor=north,xshift=1.33cm,yshift=-0.0cm}]{{$\times$ D}} & \gate{R_z(\theta_4)}   & \ctrl{1}& \qw & \qw  \\
                \qw  & \gate{H} &\gate{R_x(\theta_2)}   & \gate{R_z(\theta_5)} & \targ{} & \ctrl{1} & \qw \\
                \qw  &\gate{H} & \gate{R_x(\theta_3)} & \gate{R_z(\theta_6)}   & \qw & \targ{}   &  \qw 
\end{quantikz}}
    \caption{Ansatz 2}
  \end{subfigure}
  \caption{Two types of ansatzes used in this work. The single-qubit gate $R_i(\theta)$ represents a rotation around the $i$ axis of the Bloch sphere by an angle of $\theta$, where  $i \in \{x,y,z\}$, and $\theta \in [0,2\pi)$ is a trainable parameter. Ansatz 2 contains two-qubit CNOT gates which might create quantum entanglement. The dashed box indicates
a single circuit layer that can be repeated D times to enhance the expressive power of the ansatz.  }
\label{ansatzes}
\end{figure}

Our proposed architecture has two advantages. First, with the absence of global pooling, the complete information extracted by the backbone model is preserved and can be sufficiently exploited for classification. This is crucial for image classification tasks requiring discriminative and fine-grained details. Second, our framework generally involves fewer parameters. In VQCs, each parameterized quantum gate corresponds to one parameter, and each gate can only operate on one or two qubits. This locality helps reduce the number of parameters. In contrast, classical fully connected neural networks require each neuron to has  connection weights with all neurons in the next layer, leading to an extensive amount of parameters. Also, the number of qubits in the AE-VQC is exponentially smaller than the  size of the input features thanks to amplitude encoding, and the number of parameterized gates can be substantially reduced by designing an appropriate ansatz. For instance, in the case of 10-class classification using the ResNet-18 model, the total number of parameters after the last convolution layer is $15D + 15\times10 = 15(D+10)$ if Ansatz 1 is employed for the AE-VQC in the proposed framework. Here, $15D$ is the number of quantum parameters from the AE-VQC, where $D$ indicates the number of repeated layers in the ansatz, and $15\times10$ is the number of classical parameters from the fully connected layer.  Since choosing $D \le 10$ is sufficient to ensure the expressive power of the AE-VQC, $15(D+10)$ is generally much smaller than the number of parameters after the last convolution layer in the classical ResNet-18 model, which is $512\times 10=5120$. Furthermore, in light of the strong expressibility of the AE-VQC, our framework requires less number of fully connected layers than the classical framework for some specific tasks, which also contributes to decreasing the number of parameters.

\section{Experiments}\label{experiment}
\subsection{Datasets and Evaluation Metrics}
We evaluate the proposed approach on the following four publicly available benchmarks.
\bi
\item \textbf{Croatian Fish dataset}  includes 794 images of 12 fish species in their unconstrained natural habitat, showcasing a broad spectrum of shapes, sizes, and poses. This dataset is particularly challenging due to the high visual similarity among fish species, background clutter and varying lighting conditions. We randomly select 80\% of the data as the training set and 20\% as the test set.

\item \textbf{Aircraft dataset}  is a collection of 10,000 images of airplanes, spanning across 100 distinct aircraft models. This dataset is designed specifically for fine-grained visual classification and is a part of the ImageNet 2013 FGVC challenge. It provides a complex classification task due to subtle visual differences between aircraft models, and significant variability in aircraft design and branding. Due to resource limitations, we select a subset of the dataset, which includes 8 variants of the Boeing 737 aircraft model ranging from 737-200 to 737-900. The training and test set contain 533  and  267 images respectively.

\item \textbf{Breast Ultrasound image (BUSI) dataset}  consists of 780 breast ultrasound images among women in ages between 25 and 75 years old. This dataset is categorized into three classes: normal, benign, and malignant images. The challenge of this dataset lies in differentiating between benign and malignant tissue, a task of significant importance in the early detection of breast cancer. We find an identical image in both of the benign and malignant categories, making it impossible to determine which category these two images belong to. To avoid impacting the experiment results, we remove both images from the dataset. Finally, the training and test set include 620 and 158 images respectively. 

\item \textbf{Apples or Tomatoes (AOT) dataset}  is designed for a binary classification task which focuses on classifying images as either apples or tomatoes. Despite the apparent simplicity, the task is non-trivial due to similarities in color and shape between the two classes, especially when the fruits are partially obscured or in unusual orientations.  The dataset consists of 294 images in the training set and 97 images in the test set.

\ei
We choose accuracy and  macro-averaged F1  score over the test set as evaluation metrics to evaluate and compare our proposed framework against the classical framework.

\subsection{Training Setup}
In our experiments, we utilize the proposed framework to modify two type of classic image classification models: ResNet and MaxViT, both of which have achieved SOTA results for the task of image classification. ResNet is a family of image classification models based purely on convolutional neural networks. It is well known for the introduction of skip connections that bypass or shortcut across convolutional layers. The ResNet architecture serves as a fundamental building block in many mainstream computer vision models and represents a significant milestone in the development of deep learning. MaxViT is a family of hybrid (CNN + ViT) image classification models, that achieves better performances than both SOTA ConvNets and Transformers across a wide range of tasks such as image classification, object detection, and segmentation.  This model employs a multi-axis attention mechanism to better adapt to arbitrary input resolutions with only linear computational complexity.

Due to the resource constraints, we select lightwight ResNet-18 and MaxViT-T models from these two types of architectures as baseline models to test our proposed framework. We refer to the quantum counterparts of these two classic models as ResNet-18$\_$q and MaxViT-T$\_$q, respectively. Both ResNet-18 and MaxViT-T yield a feature map size of $512 \times 7 \times 7 = 25088$ prior to the global pooling layer. Hence in the corresponding quantum models, the number of qubits required for the AE-VQC is $\lceil \log_2 25088 \rceil  = 15$. The ansatz used in the AE-VQC varies according to the model and dataset. Detailed architectural specifications of these ansatzes are presented in Table \ref{specifications}. We train each model from scratch for 200 epochs using a mini-batch size of 32 and the Adam optimizer with a learning rate of 0.001.
\begin{table}[tbp] 
    \begin{center}
     \resizebox{0.4\textwidth}{!}{%
    \begin{tabular}{lllc} 
     \toprule
      Dataset   & Model & Ansatz & D  \\ 
      \midrule
            \multirow{3}{*}{\texttt{Croatian Fish}} &   \texttt{ResNet-18$\_$q} & Ansatz 2 & 1  \\ 
             & \texttt{MaxViT-T\_q} & Ansatz 1 & 1  \\

             \midrule
                                      \multirow{3}{*}{\texttt{Aircraft}} &   \texttt{ResNet-18$\_$q} &Ansatz 1 & 1  \\ 
             & \texttt{MaxViT-T\_q} & Ansatz 1 & 1  \\

             \midrule
                         \multirow{3}{*}{\texttt{BUSI}} &   \texttt{ResNet-18$\_$q} & Ansatz 1 & 1 \\ 
             & \texttt{MaxViT-T\_q} & Ansatz 1 & 1  \\

             \midrule

             \multirow{3}{*}{\texttt{AOT}} &   \texttt{ResNet-18$\_$q} & Ansatz 1& 1 \\ 
             & \texttt{MaxViT-T\_q} & Ansatz 1 & 3   \\

             \bottomrule
               \end{tabular}
               }
    \end{center}
    \caption{Detailed architectural specifications of the ansatzes used in the experiments. $D$ indicates the number of repeated layers in the ansatz.}
    \label{specifications}
\end{table}

\subsection{Experimental Environment}
We conduct our experiments on a local computer with a M1‌ Pro 10-core CPU by using PennyLane \cite{bergholm2018pennylane} and PyTorch \cite{paszke2019pytorch}. PennyLane is a quantum machine learning open-source library which allows for quantum differentiable programming. With a comprehensive set of features, simulators and hardware, PennyLane enables users to easily build, optimize and deploy quantum-classical applications. To build the tested models, we implement the classical modules with PyTorch and quantum modules using PennyLane. We choose the PennyLane's standard state-vector simulator  \emph{default.qubit} as the backend for executing quantum circuits in our hybrid models. This simulator supports both back-propagation \cite{Linnainmaa:1976, Rumelhart:1986we} and adjoint \cite{jones2020efficient} differentiation methods for calculating quantum gradients using the PyTorch interface.

\subsection{Results}
\begin{table}[tbp] 
    \begin{center}
     \resizebox{0.45\textwidth}{!}{%
    \begin{tabular}{llcccc} 
     \toprule
      Dataset   & Model & Params* & Acc & F1  \\ 
      \midrule
            \multirow{3}{*}{\texttt{Croatian Fish}} &   \texttt{ResNet-18} & 6,156 & 91.52\% & 91.77\% \\ 
             & \texttt{ResNet-18\_q} & 222 &\textbf{94.55}\% &\textbf{94.56}\%\\
             
                          \midrule
                         \multirow{3}{*}{\texttt{Aircraft}} &   \texttt{ResNet-18} & 4,104 & 31.46\% &31.74\% \\ 
             & \texttt{ResNet-18\_q} & 143 &\textbf{36.33}\% &\textbf{36.93}\%\\

                                      \midrule
                         \multirow{3}{*}{\texttt{BUSI}} &   \texttt{ResNet-18} & 1,539 & 80.38\% & 79.10\% \\ 
             & \texttt{ResNet-18\_q} & 63 &\textbf{86.08}\% &\textbf{84.63}\%\\

             \midrule
             
             \multirow{3}{*}{\texttt{AOT}} &   \texttt{ResNet-18}  & 1,026 & 82.47\% &82.35\%\\ 
             & \texttt{ResNet-18\_q} & 41 & \bftab 87.63\%   & \textbf{87.52}\%\\

             \bottomrule
               \end{tabular}
               }
    \end{center}
    \caption{Performance comparisons of ResNet-18 and ResNet-18\_q on four public datasets. Acc and F1 are accuracy and macro-averaged F1 score respectively reported on test sets.  Params* indicates the number of parameters following the backbone model. ResNet-18 and ResNet-18\_q have exactly the same backbone model. } 
    \label{resnet_results}
\end{table}

\begin{table}[tbp] 
    \begin{center}
     \resizebox{0.45\textwidth}{!}{%
    \begin{tabular}{llccc} 
     \toprule
      Dataset   & Model & Params* &  Acc &  F1  \\ 
      \midrule
          \multirow{2}{*}{\texttt{Croatian Fish}} &   \texttt{MaxViT-T} & 269,836 & 71.52\% &68.54\%\\ 
             & \texttt{MaxViT-T\_q} & 14,363 & \textbf{76.97}\% &\textbf{75.15}\%\\  
                 \midrule
                         \multirow{2}{*}{\texttt{Aircraft}} &   \texttt{MaxViT-T} & 267,784 & 13.11\% & 6.56\% \\ 
             & \texttt{MaxViT-T\_q} &12,311 & \textbf{22.32}\% &  \textbf{22.35}\% \\      
             \midrule           
                         \multirow{2}{*}{\texttt{BUSI}} &   \texttt{MaxViT-T} & 265,219 & 78.48\% &77.08\%\\ 
             & \texttt{MaxViT-T\_q} & 63 & \textbf{84.18}\% &\textbf{81.97}\% \\       
             \midrule    
             \multirow{2}{*}{\texttt{AOT}} &   \texttt{MaxViT-T} & 264,706 & 84.54\%& 83.98\% \\ 
             & \texttt{MaxViT-T\_q} & 107 &  \textbf{87.63}\% &  \textbf{87.56}\%\\
             \bottomrule
               \end{tabular}
               }
    \end{center}
    \caption{Performance comparisons of MaxViT-T and MaxViT-T\_q on four public datasets.}
    \label{maxvit_results}
\end{table}
We summarized the experimental results in Table \ref{resnet_results} and Table \ref{maxvit_results}. It can be observed that both quantum models outperform their classical counterparts by significant margins across all datasets, which is in consistence with our expectation. For instance, ResNet-18\_q exhibits improvements in test accuracy of up to 5.70\% and in test F1 score of up to 5.53\%, over the classical ResNet-18 model. More remarkably, MaxViT-T$\_$q provides up to 9.21\% higher accuracy and up to 15.79\% higher F1 score, when compared with MaxViT. These performance improvements mainly arise from the capability of our framework to preserve and utilize the full feature maps obtained by feature extractors of image classification models. Additionally, it is very interesting to note that Ansatz 2 generally performs worse than Ansatz 1 in our experiments, despite Ansatz 2 being more complicated and potentially capable of creating entangled states. 

Furthermore, as shown in Table \ref{resnet_results} and Table \ref{maxvit_results},  quantum models have fewer parameters than classical models, even without the use of the global pooling layer after the backbone. In particular, MaxViT-T$\_$q achieves better classification performances than MaxViT-T on BUSI and AOT Datasets, using only 0.2\% and 0.4\% of the number of parameters, respectively. This is primarily due to the powerful data storage and expressive capabilities of the variational quantum circuit with amplitude encoding. It should be noted that we only present the parameter count of the layers after the backbone model, since both the proposed and classical framework use exactly the same backbone model. 

 In our experiments, both ResNet and MaxVit are SOTA models for the task of image classification. Therefore, our experimental results not only validate the effectiveness of our approach but also demonstrate its practicality.

\section{Conclusion}\label{conclusion}
In this work, we propose a novel image classification framework based on variational quantum algorithms. The most notable feature of our framework is the elimination of the global pooling module commonly used at the end of classical image classification framework, an operation which reduces model complexity but results in a significant loss of important information. This feature enables our framework to learn discriminative details within images.  Moreover, in spite of the absence of the global pooling, our framework still has fewer parameters compared to the classical framework, preserving the capability to prevent overfitting. This stems from the adoption of the variational quantum circuit with amplitude encoding. We apply our approach to two different types of mainstream image classification models, namely the convolution-based ResNet and hybrid CNN-transformer based  MaxVit. These two benchmark models represent  state-of-the-art performances in the domain of image classification. Through extensive experiments on four challenging datasets, our proposed quantum framework exhibits superior performances over the classical framework in terms of classification metrics (e.g. accuracy and F1 score) . It is worth noting that our framework is highly flexible and can be applied to a wide range of classical image classification models.

Despite a series of achievements obtained in this work, there are still some limitations that need to be addressed in the future research. Firstly, we can explore more ansatzes for our framework  in addition to the two ansatzes used in this paper. Secondly, we perform all experiments with quantum simulators in this work. In the future, we can further evaluate our approach on real quantum hardware. Thirdly, we investigate our method only in the area of image classification. It would be worthwhile to extend our approach to other machine learning tasks such as natural language processing.

\section*{Data availability}
The data that support the findings of this study are publicly available.

\section*{Code availability}
The code that supports the findings of this study is available upon request.

\bibliographystyle{ieeetr}
\bibliography{refs}

\end{document}